\documentclass[aps,prl,a4paper,twocolumn,10pt,longbibliography,superscriptaddress]{revtex4-1}
\usepackage{amsmath}
\usepackage{amssymb}
\usepackage{mathtools}
\usepackage{graphics}
\usepackage{comment}
\usepackage{tikz}
\usepackage[normalem]{ulem}

\newcommand{\beq}{\begin{equation}}
\newcommand{\eeq}{\end{equation}}
\newcommand{\bei}{\begin{itemize}}
\newcommand{\eei}{\end{itemize}}
\newcommand{\ben}{\begin{enumerate}}
\newcommand{\een}{\end{enumerate}}

\newcommand{\be}{{\mathbf e}}

\newcommand{\br}{{\mathbf r}}

\usepackage{hyperref}
\hypersetup{
   pdfpagemode=UseNone, 
   pdfstartpage=1,
   pdfmenubar=true,
   pdftoolbar=true,
   colorlinks = true,
   linkcolor=blue,
   citecolor=blue,
   urlcolor=blue,
   bookmarksopen=false
 }

\definecolor{darkblue}{rgb}{0.,0.24,0.8}
\definecolor{britishracinggreen}{rgb}{0.0, 0.26, 0.15}
\definecolor{darkgreen}{rgb}{0,0.60,.2}

\newcommand{\pie}[1]{#1}

\def\be{\begin{equation}}
\def\ee{\end{equation}}

\begin{document}
\title{Superfluid fraction of interacting bosonic gases}
\author{Daniel Pérez-Cruz}
\affiliation{Departamento de F\'isica and IUdEA,
Universidad de La Laguna, La Laguna, 38206, Tenerife, Spain}
\author{Grigori E. Astrakharchik}
\affiliation{Departament de F\'isica, Universitat Polit\`ecnica de Catalunya, Campus Nord B4-B5, E-08034 Barcelona, Spain}
\author{Pietro Massignan}
\affiliation{Departament de F\'isica, Universitat Polit\`ecnica de Catalunya, Campus Nord B4-B5, E-08034 Barcelona, Spain}
\date{\today}
	
\begin{abstract}
The superfluid fraction $f$ of a quantum fluid is defined in terms of the response of the system to a weak and constant drag. Notably, Leggett long ago derived two simple expressions providing a rigorous upper bound and a heuristic lower bound for $f$. Here we study the superfluid fraction of bosonic gases in various two-dimensional potentials, such as regular optical lattices and disordered speckles, by solving the Gross-Pitaevskii equation and performing Diffusion Monte Carlo simulations. We show that under conditions relevant for most ultracold experiments the bounds proposed by Leggett provide a surprisingly narrow bracketing of the exact value of the superfluid fraction.
\end{abstract}
\maketitle

Superfluidity is one of the most striking effects of quantum mechanics. This phenomenon has been observed in a wide variety of systems, encompassing liquid helium, dilute Bose gases, strongly-interacting fermionic mixtures and even neutron stars~\cite{kapitza1938viscosity,Baym1969,Pethick_Smith_book_2008,Stringari_Pitaevskii_Book_2016}. In the superfluid state, a many-body ensemble is characterized by a macroscopic wavefunction, and it features remarkable properties such as frictionless flow and quantized vorticity. Generally, however, only a portion of a system is superfluid. Following Landau's two-fluid hydrodynamics~\cite{Landau1941,tisza1938transport,tisza1940theorie1,tisza1940theorie2}, the total mass density is written as $\rho=\rho_n+\rho_s$, where $\rho_n$ and $\rho_s$ are the normal and superfluid densities, respectively. At finite temperatures, a normal component arises naturally due to thermal excitations, and its interplay with the superfluid component yields second sound~\cite{Sidorenkov2013}. But a sizable $\rho_n$ can be present even at zero temperature. This happens for example in the presence of an external potential~\cite{zapata1998josephson,morsch2006dynamics,Baym2012} (the latter can even fully destroy the superfluidity, as in Mott insulators~\cite{fisher1989boson}), in spin-orbit coupled systems~\cite{martone2012anisotropic,geier2023dynamics}, and in self-organized supersolid systems  \cite{Sepulveda2010,tanzi2019observation, Blakie2024}.

In two remarkable papers, A.~Leggett noticed that a key property prompting the reduction of the superfluid fraction of a many-body system at zero temperature is the presence of density modulations (spontaneous or induced) which break translation or Galilean symmetries~\cite{leggett1970can,leggett1998superfluid}. The superfluid fraction $f=\rho_s/\rho$ is a dynamical property, but surprisingly Leggett derived simple lower and upper bounds directly from the ground state density $n=|\Psi|^2$, which is real and carries no information about the velocity field of the condensate and therefore, a priori, about dynamical properties.

Ultracold quantum gases are excellent platforms for exploring the accuracy of the Leggett's bounds. Recent works proved that the bounds are exact for dilute Bose gases in the presence of one-dimensional (1D) optical lattices~\cite{chauveau2023superfluid, tao2023observation}. Here we demonstrate by performing Diffusion Monte Carlo and Gross-Pitaevskii simulations, that quite surprisingly the bounds bracket closely the exact value of the superfluid density of dilute Bose gases in a broad class of two-dimensional (2D) regular and disordered potentials. Furthermore, we discuss the applicability of such bounds to other many-body systems.

\begin{figure}[b]
\includegraphics[width=\columnwidth]{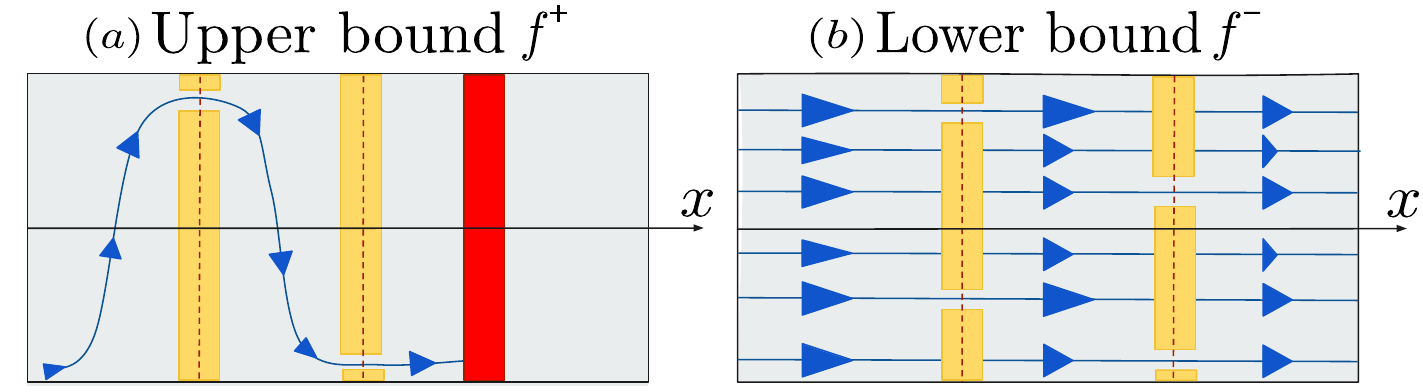}
\caption{\label{fig:intuition}
Sketches illustrating Leggett’s upper and lower bounds to the superfluid fraction $f$. 
The superflow (blue arrows) may be partially obstructed by barriers (yellow boxes), or entirely blocked by ``transverse nodal surfaces" (red box).
}
\end{figure}

{\it Superfluid fraction and Leggett's bounds.}---
Given a system of $N$ particles of mass $m$, its superfluid fraction is a tensor defined through the response along the direction $\alpha$ to a perturbation moving with velocity $v_{\beta}$ along $\beta$:
\beq
f_{\alpha\beta}=1-\lim_{v\rightarrow 0}\langle \hat{P}_\alpha\rangle/(Nmv_{\beta}),
\eeq
where $\langle \hat{P}_\alpha\rangle$ is the mean total momentum along $\alpha$. In the following, we will only consider the parallel response to a drag along $x$, and denote for simplicity $f_{xx}$ by $f$. For potentials with orthorhombic unit cell and one principal axis aligned along $x$ (or with no spatial symmetry at all), the Leggett's upper and lower bounds to $f$ read respectively~\cite{leggett1970can,leggett1998superfluid}
\begin{equation}
\label{eq:LBs}
    f^{+} = 
    \frac{1}{\bar{n}} \frac{1}{\bigl\langle \frac{1}{\langle n \rangle_{\br_\perp}}\bigl\rangle}_x,
    \;\;\;\; 
    f^{-} = 
    \frac{1}{\bar{n}} \biggl\langle \frac{1}{\langle \frac{1}{n} \rangle_x}\biggl\rangle_{\br_\perp}
\end{equation}
The upper bound is derived variationally and therefore rigorous (always exact), while the lower one is {\it heuristic} and is expected to hold only for dilute (i.e., fully-condensed) bosonic gases. Here $\bar{n} = N/L^2$ stands for the mean density, $\br_\perp$ represents all coordinates in the hyperplane perpendicular to $x$, and $\langle \cdot \rangle_\alpha$ denotes a spatial average over coordinate $\alpha =\{x,\br_\perp\}$. Whenever the particle density factorizes as $n(\br) = n_x(x)n_\perp(\br_\perp)$, the two bounds coincide:
\begin{equation}\label{eq:1DLR}
f^{+} = f^{-} = \frac{1}{\langle n_x \rangle_x\langle \frac{1}{n_x}\rangle}_x \quad\textrm{(factorized }n),
\end{equation}
and thus this expression gives the exact result for the superfluid density. This factorized form for $n(\br)$ arises for example for Bose gases in 1D optical lattices, the situation studied in the recent Refs.~\citep{chauveau2023superfluid,tao2023observation}. 

A physical intuition about these bounds may be obtained by examining the two sketches presented in Fig.~\ref{fig:intuition}. The lower bound is the easiest to understand. Consider slicing the system into 1D tubes along the axis parallel to the drag. To get $f^{-}$, one computes Eq.~\eqref{eq:1DLR} along each tube (finding $f<1$ whenever a barrier obstructs it), and averages the results over all tubes. By ignoring the cross-talk between the tubes, one obtains naturally a lower bound to $f$. The upper bound is slightly more subtle to grasp. Here one slices the system perpendicularly to the force, then extracts the superfluid fraction traversing each plane, and finally uses this quantity to compute Eq.~\eqref{eq:1DLR} along the direction of the force. By first computing the transverse average, one erases all information about the particular path the superfluid took to traverse each plane. For example, the path shown in the right panel contains sharp turns, and the superfluid is not likely to follow it because of its high kinetic cost. As a consequence, this procedure produces an upper bound $f^{+}$. Importantly, whenever the system presents {\it transverse nodal surfaces} [i.e., regions where $n(x_0,{\bf r}_\perp)=0$ for all values of ${\bf r}_\perp$, see the red box in Fig.~\ref{fig:intuition}(b)], then the upper bound immediately vanishes, and therefore the system must be fully normal.

{\it Superfluid fraction in 2D Bose gases}---In this work, we study the superfluid response of a 2D bosonic system by comparing the 2D Gross-Pitaevskii model with numerically exact Diffusion Monte Carlo (DMC) simulations of 3D Bose gases under strong transverse confinement. Our investigation covers systems in the presence of potentials which are periodic, disordered, and hybrid (ordered along one direction, and disordered along the other).
 
Our primary objective is to quantify the superfluid response of the system to a drag imparted by an external potential $V$ moving with small velocity ${\bf v}$ along $x$. 
{\bf In the Andronikashvili experiment, only the normal component of the fluid experiences friction from the rotating bucket. The normal component gets dragged by the external potential and moves with it in the lab frame.
In contrast, in the moving frame the external potential appears stationary and the normal component remains at rest relative to the potential, while only the superfluid component reacts to the drag.}
In the reference frame where the potential is stationary, the Gross-Pitaevskii equation (GPE) may be written as
\begin{equation}
\label{eq:GP-TBC}
\left[\frac{\left(-i\hbar\boldsymbol\nabla - m{\bf v}\right)^2}{2m} + V(\mathbf{r}) + g |\psi(\mathbf{r})|^2 \right]\psi(\mathbf{r}) = \mu \psi(\mathbf{r}),
\end{equation}
where $g$ is the 2D coupling constant and $\mu$ the chemical potential.  The corresponding healing length is  $\xi=\sqrt{\hbar^2/(2m\mu)}$.

\begin{figure}[b]
\includegraphics[width=\columnwidth]{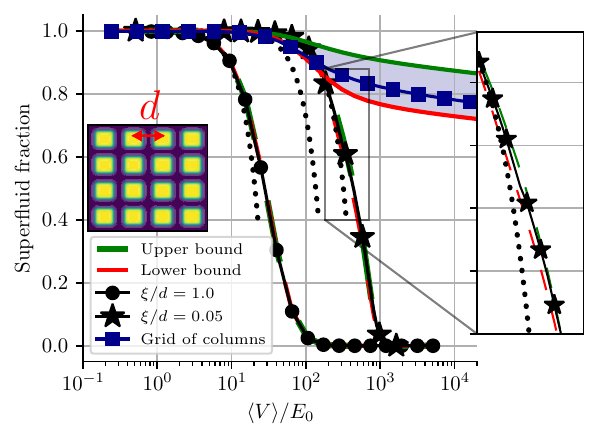}
\caption{\label{fig:opt_and_grid}
Superfluid fraction in a 2D optical lattice and in a regular grid of repulsive square columns (sketched in the inset) plotted as a function of the mean value $\langle V\rangle$ of the potential divided by $E_0 = \hbar^2/(2 m d^2)$. 
Solid symbols correspond to an optical lattice potential with lattice constant $d$ and two different values of the coupling strength. Squares show results for the column potential with $\xi/d = 0.25$. Dotted lines show the weak-perturbation Bogoliubov approximation, Eq.~\eqref{eq:fBogoliubov}.
}
\end{figure}

The perturbed and unperturbed ground state energies of Eq.~\eqref{eq:GP-TBC} are related by \cite{fisher1973helicity,PollockCeperley87}
\begin{equation}\label{eq:fTwist}
    E(v) = E(0) + f N \frac{mv^2}{2} \hspace{10pt} (v\xrightarrow{}0),
\end{equation}
an expression directly yielding the superfluid fraction $f$. This approach is particularly suitable for a numerical study because very robust numerical techniques have been developed to study the properties of the Gross-Pitaevskii equation. 
{\bf In particular, we solve Eq.~\eqref{eq:GP-TBC} using a pseudo-spectral Fourier method, in which the evaluation of the evolution operator is split between momentum and position space
(see, e.g., Refs.~\cite{mclachlan1993explicit,suzuki1991general}). 
Within this approach, the velocity-dependent term in Eq.\eqref{eq:GP-TBC} is straightforwardly implemented by means of a rigid translation by $m{\bf v}$ of all momenta, when evaluating the kinetic term in momentum space.} 

To independently assess the validity of the 2D GPE description of this system, we perform numerically exact Diffusion Monte Carlo simulations in three dimensions, adding to the potential $V$ in the $xy$ plane a tight harmonic oscillator in the $z$ direction. The interaction potential between the particles is modeled by a soft-sphere potential with the diameter equal to the transverse oscillator length $a_{\rm ho}$ and scattering length $a_{3D}$ small compared to it, $a_{3D}/a_{\rm ho}=0.1$. Under these conditions, to a good approximation, the mean-field relation between the three-dimensional parameters and two-dimensional coupling constant can be used~\cite{petrov2000bose,lee2002energy}, $g=4\pi\hbar^2 (a_{3D}/\sqrt{2\pi} a_{\rm ho})/m$. We consider $N=100$ particles in a box of size $L\times L$ with periodic boundary conditions, and we keep the 2D density $n=N/L^2$ small, so that $n a_{\rm ho}^2=0.1$. Within the MF relation for the 2D scattering length, $\xi = 3.16 a_{\rm ho}$. The guiding wave function is taken in a similar form as in Ref.~\cite{astrakharchik2013phase}, i.e., it is constructed as a product of one-body terms [Gaussian along $z$, corresponding to the ground-state of the transverse harmonic oscillator $\psi_{\text{ho}}(z)$, and GPE solution in the $(x,y)$ plane $\psi_{\perp}(x,y)$, taking into account the effect of the potential $V$] and two-body Jastrow terms (constructed by matching the solution of the two-body scattering problem at short distances and an exponential decay at large distances). 
To obtain the values of $\psi_{\perp}(x,y)$ at non-sampled points needed for the DMC integration we made use of polynomial interpolation. 

The superfluid fraction in the DMC calculations is obtained using the winding number technique~\cite{PollockCeperley87}.
The derivation is performed in the moving reference frame where the external potential is stationary, as in Eq.~\eqref{eq:GP-TBC}. By taking the limit of vanishing velocity, the superfluid density is determined from the coefficient of diffusion of the center of mass in imaginary time propagation (for a detailed derivation, see Chapter 3.3.2 in Ref.~\cite{astrakharchik2014laureathesis}). 

{\it Regular and hybrid potentials}--- In this paper we investigate the behavior of the Leggett's bounds in 2D potentials. It is important to highlight the significant conceptual difference with the case of 1D optical lattices recently considered in Refs.~\cite{chauveau2023superfluid, tao2023observation}. In the latter studies the density remained separable, and therefore by Eq.~\eqref{eq:1DLR} the bounds coincide. However, in a 2D potential the density of an interacting gas is non-separable, and therefore the accuracy of the bounds is an open question.

We start by considering two kinds of regular potentials: a symmetric 2D optical lattice with equal spatial period $d$ and intensity along both directions, and a 2D grid of square columns whose centers are separated by a distance $d$  
\cite{definitionOfColumnPotential}.
We present in Fig.~\ref{fig:opt_and_grid} our results for the superfluid fraction $f$, along with its upper and lower bounds $f^+$ and $f^-$ given by Eq.~\eqref{eq:LBs} computed over the static (${\bf v}=0$) solution of the GPE~\eqref{eq:GP-TBC}, and with the analytic Bogoliubov result valid for weak perturbations \cite{singh1994disordered,lugan2011localization,astrakharchik2013phase}
\beq\label{eq:fBogoliubov}
f_{\rm Bog}=1-\frac{2}{L^2}\int\limits_{{\bf k}\neq 0} d^2k \frac{|V({\bf k})|^2}{\left[\hbar^2k^2/(2m)+2g\bar{n}\right]^2}
\eeq
with $V({\bf k})=L^{-2}\int d^2r \,e^{-i{\bf k}\cdot{\bf r}}\, V({\bf r})$, for various values of the repulsive interactions, parametrized by means of the healing length $\xi$.

The Leggett's bounds $f^-$ and $f^+$ are seen to bracket accurately the exact value of the superfluid fraction $f$.
The bracketing is tighter for a 2D optical potential than for a grid of columns because the latter presents sharp edges. Near the edges of each column, the ground state density varies rapidly, making it harder to approximate by a factorized form, and this results in a looser bracketing.
The superfluid-to-normal transition takes place when the mean potential height becomes comparable to the chemical potential $\mu$. A notable feature visible in Fig.~\ref{fig:opt_and_grid} is that the superfluid fraction for the column potential decays very slowly for large $\langle V \rangle$. 
This is easy to understand. Our grid of columns is aligned parallel to both the $x$ and $y$ axes so that there are long unobstructed channels between them along the direction of the drag. 
In Fig.~\ref{fig:opt_and_grid} we chose $\xi=d/4$, so that the gas can easily penetrate in between the columns, no matter how tall they are. 
When they become effectively impenetrable, the gas keeps flowing around the columns (the slow decay of $f$ at large $\langle V \rangle$ is due to the exponentially small but non-zero height of the pillars across the channels \cite{definitionOfColumnPotential}).

\begin{figure}[b]
\includegraphics[width=\columnwidth]{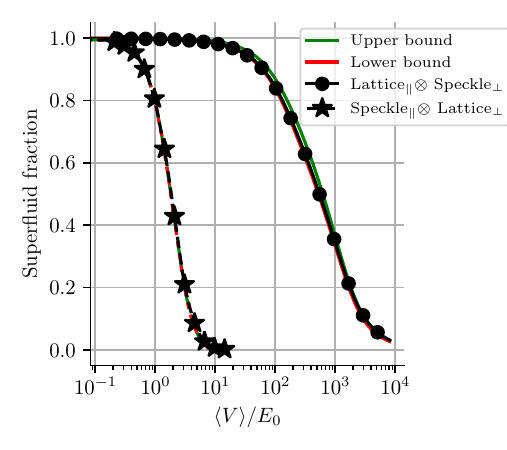}
\caption{\label{fig:mixed_potential}
Superfluid fraction in a hybrid potential. Stars denote disordered potential (speckles) along the direction of the drag and ordered (optical lattice) along it, while filled circles denote the opposite situation. 
The results are for $\xi/L_c = 1$, and averaged over five different speckle realizations.
}
\end{figure}

We continue by studying the superfluid fraction in a hybrid potential, obtained by summing 1D disordered speckles along one direction and a 1D optical lattice along the other. This asymmetric potential is illustrative because rotating it by 90$^\circ$ allows us to probe directly the tensor nature of the superfluid fraction \footnote{Here we consider time-reversal symmetric Hamiltonians and potentials with one symmetry axis aligned along the direction of the drag (or no spatial symmetry at all). In this case, the off-diagonal elements of the superfluid fraction tensor vanish, and the latter is more simply a vector.}. Speckle patterns are obtained by diffracting light off a rough surface, and constitute the simplest form of disorder since they are characterized by a single dimensionless parameter, the ratio of their mean amplitude $\langle V\rangle$ to the characteristic energy $E_0 = \hbar^2/(2 m L^2_c)$ obtained from the correlation length $L_c$ of the speckles~\cite{clement2006experimental}. We generate these patterns by computing the Fourier transform of aperture functions whose entries are random complex numbers with unit modulus, thereby mimicking the random and uncorrelated scattering events on the rough surface. We consider the simplest situation where the spatial period $d$ of the optical lattice coincides with the correlation length $L_c$ of the speckles, and the mean amplitude equals $\langle V \rangle$ in both directions. 

As shown in Fig.~\ref{fig:mixed_potential}, radically different behaviors are observed when the drag is applied parallel to the disordered direction as opposed to the periodic one. The first case closely resembles the idealization used to discuss the lower bound at the beginning of this work.  The optical lattice ``slices" the system into 1D tubes along the axis of the drag, with limited cross-talk between them, and all tubes contain the same speckle potential. The dynamics is then basically 1D, and as soon as the disorder contains a speckle which is sensibly higher than the chemical potential, all tubes are suddenly blocked (i.e., there appears a transverse nodal surface), and the superfluid density vanishes rapidly through a rather abrupt transition. The second case is very different: now we have tubes with very variable cross-talk, and along each tube the fluid sees a periodic potential. In this situation, it is much harder to form transverse nodal surfaces, and the resulting superfluid-normal transition is smoother and takes place at much larger values of $\langle V \rangle$.

\begin{figure}[b]
\includegraphics[width=\columnwidth]{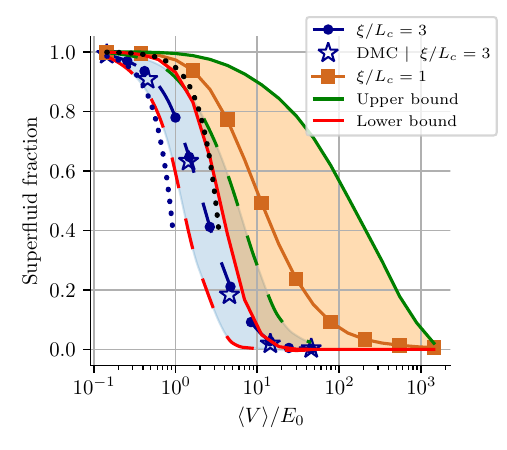}
\caption{\label{fig:bounds_speck}
Superfluid fraction in 2D speckle disorder, plotted against the mean height $\langle V\rangle$ of the speckles.
Dashed/solid curves with filled symbols represent the GPE results for a gas with weaker/stronger repulsive interactions obtained using Eq.~\eqref{eq:fTwist}. The open stars denote DMC data, the dotted lines are the weak-perturbation Bogoliubov approximation Eq.~\eqref{eq:fBogoliubov}, and the shaded regions show the windows left open by Leggett's lower and upper bounds $f^-$ and $f^+$ (computed from the ground states of the GPE). Results averaged over five speckle realizations. The statistical errors (arising from both the DMC method and the averaging over disorder realizations) are smaller than the symbols' size. No significant difference between GPE and DMC was found for the bounds.
}
\end{figure}

{\it 2D speckle disorder}---We conclude our analysis by studying the superfluidity in 2D speckle potentials  \cite{astrakharchik2013phase,damski2003anderson}. Our results for the superfluid fraction $f$ and its bounds $f^{-}$ and $f^{+}$ are shown in Fig.~\ref{fig:bounds_speck}. Finite-size effects are important for 2D disorder, and we account for them by doing calculations for increasing system sizes $L$, and extrapolating to the infinite size limit \footnote{In agreement with  Ref.~\cite{astrakharchik2013phase}, we find that the superfluid fraction scales linearly with the inverse system size: $f(\infty) = f(L) +c/L +\ldots$, with $c<0$. }. In analogy with the cases studied before, the superfluid fraction starts to decrease when the mean disorder amplitude exceeds the chemical potential, and it becomes significantly suppressed when $\langle V\rangle\gg \mu$. The GPE compares very favorably with exact DMC when the gas is sufficiently dilute, a condition which is well-verified for $\xi/L_c=3$.

Leggett's bounds are seen to bracket correctly the superfluid density, and the window between them is particularly narrow for weak repulsive interactions, such that $\xi>L_c$. Under these conditions, the superfluid-insulator transition happens for very weak potentials, so that the ground state wavefunction is approximately separable, and Eq.~\eqref{eq:1DLR} shows that the two bounds should be very close to the actual value of $f$. On the other hand, for strong disorder (large $\langle V \rangle)$ the superfluid fraction and its bounds all tend rapidly to zero, signaling the appearance of transverse nodal surfaces in the ground state density.

{\it Discussion and conclusions} --- We have shown that the bounds derived by Leggett bracket with surprising accuracy the superfluid fraction of dilute Bose gases, which can be described by the Gross-Pitaevskii equation, under many experimentally-relevant situations. Our findings complement previous studies where the bounds were used to study Josephson junctions~\cite{zapata1998josephson, Pezze2023,biagioni2023sub}.

When Leggett first derived his upper bound for the superfluid fraction, he was discussing the viability of a supersolid phase in crystalline systems~\cite{leggett1970can}. Using this bound he estimated that the superfluid fraction in solid helium should be very small (``probably $\leq 10^{-4}$").
On the other hand, supersolidity has been recently observed in spin-orbit coupled BECs, BECs in cavities, dipolar quantum gases and dilute Bose-Bose mixtures \cite{leonard2017supersolid,li2017stripe,tanzi2019observation,bottcher2019transient,Chomaz2019,Recati2023,Ripley2023,Sindik2023,Hirthe2024}. There, density modulations appear spontaneously in the so-called ``stripe-phase", under conditions where the mean-field description remains accurate. As such, we expect that the lower Leggett's bound will apply, and that the two bounds will accurately bracket the actual value of $f$.

Correctly understanding when the Leggett's bounds are applicable and when they are accurate is of great interest to the cold atom community, since these bounds provide a direct way to estimate $f$ by the sole {\it in-situ} observation of the particle density.
On the other hand, to measure the actual $f$ one would need to read out the response of the system to an infinitesimal perturbation, which is often experimentally impractical. 

{\bf Nonetheless, the picture of superfluid flow underlying the derivation of the lower Leggett's bound is essentially a classical (mean-field) one and does not take into account interference and quantum correlations, which have substantial effects in specific cases.}
For example, there are various situations where the lower bound (which is heuristic, i.e., not rigorous) fails. First, this happens in systems possessing gapless excitations at finite momentum. In such cases, the Landau criterion predicts that superfluidity is strongly reduced~\cite{Baym2012}. At the same time, the energy might remain quadratic in the phase twist and the system could have a uniform density, so that both the lower Leggett's bound $f^{-}$~\eqref{eq:LBs} and the definition of $f$ in Eq.~\eqref{eq:fTwist} would predict a fully superfluid system. This is the case for example in an ideal Bose gas (which is pathologic as it has a vanishing sound speed), but also in ideal Fermi gases in any dimensionality or in 1D Luttinger liquids, which possess a non-zero sound speed, but have gapless Umklapp processes at twice the Fermi momentum~\cite{Cazalilla_2004, astrakharchik2004motion, carusotto2004superfluidity}.

Similarly, $f^{-}$ fails for systems which are unstable to infinitesimal perturbations, such as the ideal Fermi gas (which in this case displays an Anderson orthogonality catastrophe) or Luttinger liquids with Luttinger parameter $K<3/2$ in presence of disorder~\cite{Giamarchi1998,Giamarchi04,cazalilla2011one}. These systems retain an almost uniform density (so that $f^{-}\approx 1$) even if their energy is unaffected by the drag (and thereby one must have $f=0$).

The upper bound $f^{+}$ is instead rigorous, and therefore applies to generic many-body systems. For example, Orso and Stringari proved by means of sum rules that $f^{+}$ accurately approximates $f$ in a unitary two-component Fermi gas, provided that the external perturbation probes the phononic regime of small momentum, i.e., the paired nature of the system \cite{Orso2023}. At higher momenta, however, the spectrum is no longer simply phononic, and $f^{+}$ overestimates $f$.

Interesting open directions for future work include studying the accuracy of Leggett's bounds for strongly-interacting bosons~\cite{Yao2023}, and their extension to non-zero temperatures.

\vspace{5mm}
\begin{acknowledgments}
We wish to thank David Clément, Jean Dalibard, Laurent Sanchez-Palencia and Sandro Stringari 
for insightful discussions.
We acknowledge support by the Spanish Ministry of Science and Innovation (MCIN/AEI/10.13039/501100011033, Grants No. PID2020-113565GB-C21 and PID2023-147469NB-C21), by the Spanish Ministry of University (grant FPU No. FPU22/03376 funded by MICIU/AEI/10.13039/501100011033), and by the Generalitat de Catalunya (grant 2021 SGR 01411).
P.M.~further acknowledges the {\it ICREA Academia} program, the Institut Henri Poincaré (UAR 839 CNRS-Sorbonne Université) and the LabEx CARMIN (ANR-10-LABX-59-01).
\end{acknowledgments}

\bibliography{BIBLIOGRAPHY}

\begin{thebibliography}{55}
\bibitem [{def()}]{definitionOfColumnPotential}%
  \BibitemOpen
  \href@noop {} {}\bibinfo {note} {\pie{The square grid of columns is defined as $\propto \sum_{i,j}V_{\rm col}^{(i)}(x)\cdot V_{\rm col}^{(j)}(y)$; along $x$ (and similarly along $y$) each pillar is modeled as $V_{\text{col}}^{(i)}(x) = \tanh{\left[16 \left(x+i+l/2\right)/d\right]}-\tanh{\left[16\left(x+i-l/2\right)/d\right]}$, with the integer $i$ fixing its center and $l\approx d/2$ determining its linear size.}}\BibitemShut {Stop}%
\end{thebibliography}

\end{document}